%% file: conference_101719.tex
\documentclass[10pt, conference, letterpaper]{IEEEtran}
\IEEEoverridecommandlockouts

\usepackage{cite}
\usepackage{amsmath,amssymb,amsfonts}

\usepackage{algorithmic}
\usepackage{graphicx}
\usepackage{textcomp}

\usepackage[ruled,vlined,linesnumbered]{algorithm2e}
\usepackage{subcaption}
\usepackage[hyphens]{url}
\usepackage[breaklinks=true]{hyperref}
\hypersetup{hidelinks=true, }
\usepackage[nameinlink]{cleveref}
\usepackage{fontawesome5}
\usepackage{stmaryrd}
\usepackage{wrapfig}
\usepackage{fontawesome5}

\usepackage{cancel}

\usepackage{multirow}
\usepackage[table,xcdraw]{xcolor}
\usepackage[normalem]{ulem}

\newboolean{showcomments}
\setboolean{showcomments}{true}
\ifthenelse{\boolean{showcomments}}
{ \newcommand{\mynote}[3]{
        \fbox{\bfseries\sffamily\scriptsize#1}
        {\small\textsf{\emph{\color{#3}{#2}}}}}}
{ \newcommand{\mynote}[3]{}}

\useunder{\uline}{\ul}{}
\def\BibTeX{{\rm B\kern-.05em{\sc i\kern-.025em b}\kern-.08em
    T\kern-.1667em\lower.7ex\hbox{E}\kern-.125emX}}
\begin{document}

\newcommand{\sysname}{TriHaRd\xspace}

\title{\sysname: Higher Resilience for TEE Trusted Time
\thanks{This work was supported by a French government grant managed by the Agence Nationale de la Recherche for the STEEL project of the CLOUD PEPR under the France 2030 program, reference ``ANR-23-PECL-0007'', as well as the ANR Labcom program, reference ``ANR-21-LCV1-0012''.}
}

\newboolean{anon}
\setboolean{anon}{false}
\newboolean{oneline}
\setboolean{oneline}{true}
\ifthenelse{\boolean{anon}}{
    \author{\IEEEauthorblockN{Anonymous Authors}
    Submission ID: 1571182596
    }
}
{
    \ifthenelse{\boolean{oneline}}{
        \author{
            \IEEEauthorblockN{Matthieu Bettinger\text{*}\textsuperscript{§}, Sonia Ben Mokhtar\text{*}, Pascal Felber\textsuperscript{°},\\ Etienne Rivière\textsuperscript{$\ddagger$}, Valerio Schiavoni\textsuperscript{°}, Anthony Simonet-Boulogne\textsuperscript{\textdagger}}
            \IEEEauthorblockA{\text{*}INSA Lyon, CNRS, Universite Claude Bernard Lyon 1, LIRIS, UMR5205, 69621 Villeurbanne, France\\
            \{given-name\}.\{surname\}@liris.cnrs.fr \textsuperscript{§}Corresponding author}
            \IEEEauthorblockA{\textsuperscript{°}University of Neuchâtel, 2000 Neuchâtel, Switzerland
            \{given-name\}.\{surname\}@unine.ch}
            \IEEEauthorblockA{\textsuperscript{$\ddagger$}ICTEAM, UCLouvain, 1348 Louvain-la-Neuve, Belgium
            \{given-name\}.\{surname\}@uclouvain.be}
            \IEEEauthorblockA{\textsuperscript{\textdagger}iExec Blockchain Tech, 69008 Lyon, France
            \{given-name\}.\{surname\}@iex.ec}
            Accepted for publication in the 2026 45th IEEE International Conference on Computer Communications (INFOCOM)\\
            (DOI pending)
            \vspace{-0.25cm}
            }
    }
    {
        \author{\IEEEauthorblockN{Matthieu Bettinger}
        \IEEEauthorblockA{\textit{LIRIS-DRIM INSA Lyon} \\
        Lyon, France \\
        matthieu.bettinger@liris.cnrs.fr}
        \and
        \IEEEauthorblockN{Sonia Ben Mokhtar}
        \IEEEauthorblockA{\textit{LIRIS-DRIM CNRS} \\
        Lyon, France \\
        sonia.ben-mokhtar@liris.cnrs.fr}
        \and
        \IEEEauthorblockN{Pascal Felber}
        \IEEEauthorblockA{\textit{University of Neuchâtel} \\
        Neuchâtel, Switzerland \\
        pascal.felber@unine.ch}
        \and
        \IEEEauthorblockN{Etienne Rivière}
        \IEEEauthorblockA{\textit{ICTEAM, UCLouvain} \\
        Louvain-la-Neuve, Belgium \\
        etienne.riviere@uclouvain.be}
        \and
        \IEEEauthorblockN{Valerio Schiavoni}
        \IEEEauthorblockA{\textit{University of Neuchâtel} \\
        Neuchâtel, Switzerland \\
        valerio.schiavoni@unine.ch}
        \and
        \IEEEauthorblockN{Anthony Simonet-Boulogne}
        \IEEEauthorblockA{\textit{iExec Blockchain Tech} \\
        Lyon, France \\
        anthony.simonet-boulogne@iex.ec}
        }
    }   
}

\maketitle


\begin{abstract}
    Accurately measuring time passing is critical for many applications.
    However, in Trusted Execution Environments (TEEs) such as Intel SGX, the time source is outside the Trusted Computing Base: a malicious host can manipulate the TEE's notion of time, jumping in time or affecting perceived time speed. 
    Previous work (Triad) proposes protocols for TEEs to maintain a trustworthy time source by building a cluster of TEEs that collaborate with each other and with a remote Time Authority to maintain a continuous notion of passing time.
    However, such approaches still allow an attacker to control the operating system and arbitrarily manipulate their own TEE's perceived clock speed.
    An attacker can even propagate faster passage of time to honest machines participating in Triad's trusted time protocol, causing them to skip to timestamps arbitrarily far in the future.
    We propose \sysname, a TEE trusted time protocol achieving high resilience against clock speed and offset manipulations, notably through Byzantine-resilient clock updates and consistency checks.
    We empirically show that \sysname~mitigates known attacks against Triad. 
\end{abstract}


\input{content/introduction}
\input{content/related_work}
\input{content/problem_statement}
\input{content/solution}
\input{content/experimental_protocol}
\input{content/results}
\input{content/conclusion}
\input{content/bibliography}
\end{document}

%% file: content/introduction.tex
\section{Introduction}

While Trusted Execution Environments (TEEs) such as Intel SGX enforce integrity and confidentiality, applications relying on them still suffer from timing attacks. 
These are attacks in which time is manipulated by an adversary, making timestamps unreliable~\cite{FinkenzellerBRH24}. 
The impacts of timing attacks are diverse and may have harmful consequences. 
Examples include circumventing credential expiration~\cite{Alder_Scopelliti_VanBulck_Muhlberg_2023,Malhotra_Cohen_Brakke_Goldberg_2015}, cheating in latency-sensitive systems such as auctions~\cite{Addison_Andrews_Azad_Bardsley_Bauman_Diaz_Didik_Fazliddin_Gromoa_Krish_etal_2019}, disturbing the consistency of distributed databases~\cite{Corbett_Dean_Epstein_Fikes_Frost_Furman_Ghemawat_Gubarev_Heiser_Hochschild_etal_2013}, decorrelating data and timestamps in time-series~\cite{nasrullahTrustedTimingServices2024}, or re-ordering requests to favor colluding clients in a decentralized marketplace~\cite{bettinger2025cooltee}.

The main reason why CPU-level TEEs are prone to timing attacks is that the time source is outside the Trusted Computing Base (TCB), e.g., the enclaves in Intel SGX~\cite{Costan_Devadas_2016} terminology.
In response to the risks of timing attacks, in recent years, new protocols have been proposed to address trusted time in Intel SGX, including T3E~\cite{hamidyT3EPracticalSolution2023a} and Triad~\cite{fernandezTriadTrustedTimestamps2023}.
T3E uses a Trusted Platform Module (TPM) as a trustworthy time source and implements timestamps with a predefined, limited number of usages.
Attacks delaying messages from the TPM to the application consuming timestamps cause a drop in throughput, argued to be detectable by a client~\cite{hamidyT3EPracticalSolution2023a}.
However, configuring the proper number of uses per timestamp is difficult in practice as applications may consume timestamps at different rates over different sections of their code.
The necessary rate of timestamp acccesses may, moreover, depend on the underlying hardware's performance as well as on the request load.
The former is further possibly impacted by the number of remote clients, further complicating attack detection.
Removing the burden of attack detection from clients, Triad~\cite{fernandezTriadTrustedTimestamps2023} proposes a distributed time service where a cluster of TEEs collaborates to keep a shared notion of trusted time, assuming that all underlying OSs or hypervisors may be compromised.
Triad combines local detection of interruptions in the enclave's execution by the OS with a remote protocol to recover a new time reference, either from peer enclaves in the TEE cluster or from a remote Time Authority (TA).
Unfortunately, Triad is vulnerable to a single compromised node that can manipulate its own notion of time by delaying some protocol messages with the TA~\cite{bettinger2025trihard}.
In subsequent communications, the compromised node is able to cause honest nodes to skip to timestamps arbitrarily far in the future.
This effect then cascades to other honest nodes contacting infected honest nodes, infecting them in turn.

In this paper, we propose \sysname, a TEE-based trusted time protocol with high resilience against attacks manipulating enclave-perceived clock speeds and offsets.
We contribute local and remote protocols to be resilient to adversaries controlling up to a half minus one of the hosts' OSs in the cluster.
\sysname uses optimistic synchronization to a TA (e.g., an NTP time server), allowing honest nodes to obtain accurate timestamps.
Additionally, fine-grained clock monitoring is performed locally on the machine and between TEE cluster peers to reduce the request load on the TA.
Finally, in contrast with Triad~\cite{fernandezTriadTrustedTimestamps2023}, communications between peers do not cause clock parameter updates, denying an attack vector to malicious machines, while still enabling clock consistency checks.
Malicious nodes' enclaves failing synchronization or clock checks are prevented from serving timestamps to client applications. 
Using single- and multi-node deployments, we show how \sysname mitigates attacks on Triad~\cite{bettinger2025trihard}. 
\sysname additionally reduces local clock drift to be within standard tolerance bounds, e.g., NTP~\cite{NetworkTimeProtocol_1985}'s $15\mu s/s$.
We support experimental reproducibility and make our artifact openly available~\cite{TriHaRdcode}.

This paper is organized as follows.
\Cref{sec:related_work} presents related work on TEE trusted time mechanisms.
\Cref{sec:problem_statement} describes the considered system and attacker models for TEE trusted time.
\Cref{sec:solution} details the proposed \sysname protocol and how it protects against attacks on the TEEs' perception of time.
\Cref{sec:experimental_protocol} describes our experimental protocol, and \Cref{sec:results} empirically evaluates \sysname in fault-free and adversarial setups, comparing attack impacts to those with Triad.
Finally, \Cref{sec:conclusion} summarizes this work.

%% file: content/related_work.tex
\section{Related work}\label{sec:related_work}

Trusted time mechanisms in TEEs rely on time sources such as the TimeStamp Counter (TSC) register on the CPU or remote Time Authorities (TA, e.g., NTP time servers~\cite{NetworkTimeProtocol_1985,NTPsec}), which lie outside the TCB. 
This makes them susceptible to tampering by attackers operating at the hypervisor, OS, or network level. 
To address this, security measures are essential to safeguard timestamps from manipulation. 
Alder et al.~\cite{Alder_Scopelliti_VanBulck_Muhlberg_2023} categorize security levels for trusted time in TEEs. 
The following sections discuss recent work and its characteristics.


\subsection{TimeStamp Counter (TSC) and virtualization}

In recent Intel processor hardware~\cite{intelManual}, the value in the TSC register increments at a fixed rate, such that it can be used as a wall clock, i.e., it is independent of the frequency of CPU cores.
However, privileged software like the operating system can modify the value in the TSC's Model Specific Register (MSR).
This MSR value is separate for each logical core (e.g., in case of hyper-threading).
Additionally, in the case of virtualization, a TSC offset and multiplier are available to the hypervisor to adapt read TSC values for each VM, e.g., after migration between cores.
VM execution controls define whether the TSC is virtualized or directly accessed, as well as whether its value is offset and scaled.

\subsection{CPU-level TEE trusted time}

With the first generation of Intel SGX hardware (SGX1), the enclave cannot access the TSC directly: it must delegate the call to untrusted code, enabling an attacker to modify the value. 
In contrast, the more recent Scalable Intel SGX (SGX2) permits enclaves to execute the \texttt{rdtsc} instruction, thereby bypassing the OS.
The enclave can access the raw value of the TSC if it is not virtualized by the hypervisor and not offset or scaled.
Even so, the operating system retains control over process scheduling and can, therefore, interrupt enclave execution at will. 
During interruptions, the OS or hypervisor can modify the TSC value seen by the core running the enclave.
AEX-Notify~\cite{Constable_Bulck_Cheng_Xiao_Xing_Alexandrovich_Kim_Piessens_Vij_Silberstein_2023} provides a mechanism for Intel SGX enclaves to detect Asynchronous Enclave eXits (AEXs, i.e., enclave interruptions), and respond to them.
Namely, it executes developer-defined code once the enclave resumes.

Triad~\cite{fernandezTriadTrustedTimestamps2023} proposes a distributed approach where a cluster of TEEs collaborates to maintain a continuous and shared notion of time. 
Each TEE monitors its TSC and uses AEX-Notify to detect interruptions that disrupt monitoring continuity. 
When such disruptions occur, the TEE either retrieves a timestamp from a peer enclave in the cluster or, as a fallback, consults a TA.
However, vulnerabilities in Triad's calibration protocol and inter-TEE communication have been identified~\cite{bettinger2025trihard}, enabling time manipulations at individual nodes, which can then propagate across the cluster.
Indeed, clock speed calibration is based on short-duration measurements in communications with the TA.
An attacker selectively adding delays to protocol messages can manipulate the estimated TSC frequency.
Moreover, peers in the TEE cluster update their clocks based on each other, using the highest-valued clock: a miscalibrated fast clock causes all clocks to drift toward it.

T3E~\cite{hamidyT3EPracticalSolution2023a} leverages a TPM as a colocated time source for the TEE. 
Depending on the TPM implementation, it takes tens to hundreds of milliseconds to obtain verified timestamps, so T3E allows timestamps to be used multiple times. 
To mitigate message delay attacks, T3E limits the number of times a single timestamp can be reused by the TEE and stalls execution when this limit is exceeded. 
However, determining optimal usage limits is challenging, as they depend on the specific code, workload, and hardware. 
Moreover, monitoring throughput becomes complex in scenarios with non-interactive applications or multiple users, some of whom may be malicious. 
The TPM itself is not immune to attacks, as it can be misconfigured by an attacker, leading to significant drift rates (up to $\pm32.5\%$ compared to real time~\cite{TPM20Library}).

\subsection{VM-level TEE trusted time}

VM-level TEEs, such as Intel TDX and AMD SEV-SNP, are increasingly adopted by Cloud Service Providers like Microsoft Azure~\cite{MicrosoftAzure2023}, Google Cloud~\cite{IntelGoogleCloud_2024}, and IBM Cloud~\cite{RunIBMenclave_2025}. 
These TEEs introduce mechanisms to protect their time sources even against a malicious hypervisor. 
For instance, Intel TDX employs a virtualized TSC~\cite{tdxspecs2023}, which prohibits modifications to the TSC's registers from within the Trust Domain and detects hypervisor-induced offsets during VM exits, triggering an error upon re-entry~\cite{tdxspecs2023}. 
Similarly, AMD SEV-SNP's SecureTSC feature ensures that TSC modifications by the hypervisor or guest VMs do not affect other guests, maintaining a linearly increasing TSC for each guest~\cite{Neela}.

Despite these improvements, VM-level TEEs present a larger TCB, which is more challenging to audit: ongoing research explores their attack surface~\cite{Neela,Gast_Weissteiner_Schröder_Gruss_2025,Wilke_Wichelmann_Rabich_Eisenbarth_2023,Mandal_Shukla_Mishra_Bhattacharya_Saxena_Mukhopadhyay_2025,Wilke_Sieck_Eisenbarth_2024}. 
In this paper, we aim to achieve similar trusted time guarantees as VM-level TEEs but leverage CPU-level TEEs with a smaller TCB.


%% file: content/problem_statement.tex
\section{Problem statement}\label{sec:problem_statement}

In this section, we describe the model of TEE time services for client applications, 
as well as the model of an attacker trying to impact client application behavior via attacks against the time service they consume, illustrated by \Cref{fig:system-overview}. 

\subsection{Distributed TEE time service model}

We consider $N$ SGX2-enabled nodes in a cluster, for example, in the same cloud datacenter. 
These nodes host client applications, running in enclaves and requiring on clock synchronization between nodes, e.g., within an offset of 1ms.
Additionally, an enclave application on each node runs a time service, from which client applications can get the local clock's current timestamp.
An available source of trust for time reference and speed is a remote TA node with authentication capabilities, e.g., an NTS server~\cite{Franke_Sibold_Teichel_Dansarie_Sundblad_2020}.

\subsection{Threat model}

As for previous TEE Trusted Time protocols presented in the literature~\cite{hamidyT3EPracticalSolution2023a,fernandezTriadTrustedTimestamps2023}, we assume the attacker wants to preserve the availability of the time service to client applications.
Client applications should not be able to discriminate between attacked and benign time services.
The attacker's objective is to manipulate the notion of time of client applications via attacks on the time service's clock, e.g., changing the speed or time reference, making forward or backward jumps in time.

We consider an attacker positioned on at most $f$ malicious nodes among the $N=2f+1$ total nodes in the cluster.
Indeed, given that time services run in enclaves, the attacker cannot forge messages nor equivocate.
Similarly to other TEE-based Byzantine Fault Tolerance protocols~\cite{Wang_Deng_Niu_Reiter_Zhang_2022,Stathakopoulou_Rusch_Brandenburger_Vukolic_2021,Zhang_Gao_Wang_Wu_Li_Guan_Chen_2022,Behl_Distler_Kapitza_2017,dinis2023rr,Gao_Dang_Chang_Li_2022,Messadi_Becker_Bleeke_Jehl_Mokhtar_Kapitza_2022}, TEE usage raises tolerance to malicious nodes from a constraint of $N\geq3f+1$ to $N\geq2f+1$.
Still, the attacker can try to leverage communications between malicious nodes and honest ones to propagate attacks~\cite{bettinger2025trihard}. 
The attacker can also try to hide local manipulations from other nodes, e.g., by having a slow clock but jumping to the correct time when necessary.

To do so, the attacker controls the OS and possibly the hypervisor on top of which the time service and client application enclaves run.
In particular, the attacker can manipulate process scheduling, cause interruptions to a running process, and delay or drop messages on the network to and from enclaves.
For example, the attacker can exploit the network delay symmetry assumption~\cite{FinkenzellerBRH24} in time protocols like NTP~\cite{NetworkTimeProtocol_1985} and PTP~\cite{9120376}: adding a one-way delay either to the request or response message in those protocols makes the machine perceive its own clock to have an offset to the past or the future, respectively.
If the service runs inside a VM, we assume \texttt{rdtscp} instructions to read the TSC register do not cause a VM exit, i.e., ``RDTSC exiting'' VM-execution control is 0~\cite{intelManual}.
With that configuration, enclave execution is not interrupted when the TSC is accessed. 
In turn, as long as there is no interruption while in enclave mode, the attacker cannot alter the TSC value.
However, the TSC offset and multiplier may be arbitrarily manipulated when a given core is not in enclave mode.

Using those attack levers to change the time service's perceived clock speed or offset, the attacker impacts timestamps served to client applications. 
Due to the extensive and often critical use of timestamps, it enables breaking clock accuracy assumptions in a wide array of protocols.
This applies both to TEE-based and traditional systems, notably impacting security in network communications~\cite{Malhotra_Cohen_Brakke_Goldberg_2015}, resource allocation and billing in Cloud~\cite{trachTLeaseTrustedLease2021, 10.1145/3338466.3358916}, machine-learning accuracy in IoT~\cite{nasrullahTrustedTimingServices2024}, consistency in distributed databases~\cite{Corbett_Dean_Epstein_Fikes_Frost_Furman_Ghemawat_Gubarev_Heiser_Hochschild_etal_2013} and quality-of-service in search engines~\cite{bettinger2025cooltee}.

%% file: content/solution.tex
\section{Solution}\label{sec:solution}

\begin{figure}
    \includegraphics[trim={1mm 0mm 1mm 1mm}, scale=1]{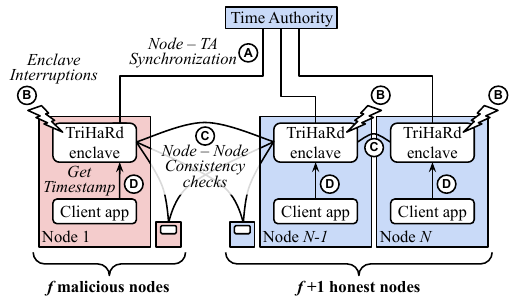}
    \caption{\sysname distributed time service overview.}
    \label{fig:system-overview}
    \vspace{-0.33cm}
\end{figure}

\begin{figure}
    \includegraphics[trim={1mm 0mm 1mm 1mm}, scale=0.5]{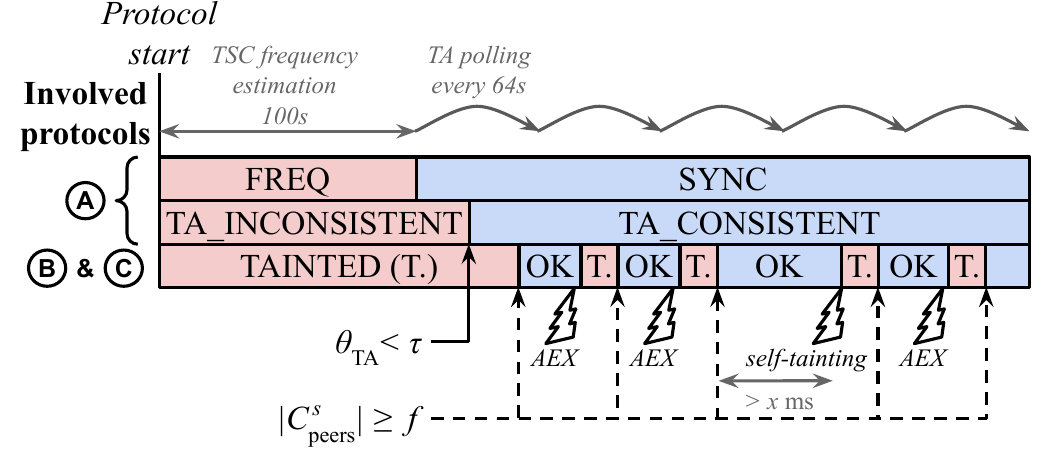}
    \caption{Example evolution of \sysname node states over time (not to scale), as well as events and conditions triggering state changes. 
    Protocols that update given states are indicated on the left.
    States in red prevent getting timestamps in protocol \textbf{D}, while states in blue are necessary. 
    $\tau$ is the maximum absolute TEE--TA drift; 
    $|C^{s}_{\text{peers}}|$ is the size of the set of consistent peers for protocol \textbf{C}'s message sequence number $s$.
    $x$ is the maximum time the enclave may remain in the \texttt{OK} state.
    States notably satisfy the constraints: $\texttt{OK}\implies\texttt{TA\_CONSISTENT}\implies\texttt{SYNC}$, as well as $\texttt{FREQ}\implies\texttt{TA\_INCONSISTENT}\implies\texttt{TAINTED}$.
    }
    \label{fig:system-example}
    \vspace{-0.33cm}
\end{figure}

\sysname comprises four sub-protocols to enable a resilient timestamp service for client applications, as depicted in \Cref{fig:system-overview}.
\textbf{A)} Each \sysname enclave performs synchronization with the TA, in this work, using a simplified NTP protocol.
\textbf{B)} An enclave thread monitors the TSC over time and stores the current value. 
The stored TSC value is invalidated after interruptions occur. 
During the interruption, the attacker may, indeed, have manipulated the TSC.
\textbf{C)} To verify the TSC value while easing the request load on the TA, enclaves in the cluster cooperate to check whether their clocks (i.e., TSC values and reference timestamps) are consistent.
\textbf{D)} Finally, a client enclave application can get the current timestamp from the \sysname enclave, after some checks on the stored TSC value's state.
All communications with enclave client applications, enclave peers in the cluster, and the TA are authenticated, respectively, using local-, remote-TEE attestation, and the TA's certificate.
In the following sections, we detail the sub-protocol with the matching letter. 
\Cref{fig:system-example} displays a time diagram of an example evolution of \sysname node states described hereafter.

\subsection{Clock synchronization}

First, the \sysname enclave clocks on all $N$ nodes should calibrate their speed and reference time.
Protocol~\textbf{A} is designed for honest nodes to obtain well-synchronized clocks, while subsequent protocols ensure resilience to attacks.
In this work, a simplified NTP protocol is used.
To synchronize with a TA~\cite{NetworkTimeProtocol_1985}, an enclave sends a message containing its time of emission ($T_{1}$) using the local clock.
The TA responds with its own message reception ($T_{2}$) and retransmission ($T_{3}$) times, while the enclave logs the TA response's reception time ($T_{4}$).
Given these four timestamps, a perceived clock offset can be computed as $\theta=\frac{T_{2}-T_{1}+T_{3}-T_{4}}{2}$.
These measurements are used in the NTP protocol to adjust the clock's offset and speed. 

Enclaves start with NTP's \texttt{FREQ} phase, where they measure an increase in TSC value between two messages from the TA, with a large duration separating them (e.g., 100s).
At the end of this phase, the TSC frequency with respect to this reference duration is computed and stored.
The node then switches to NTP's \texttt{SYNC} phase: a controller disciplines the local clock to the TA's reference with small updates in phase and frequency.
Clock updates have a small magnitude: honest nodes only need to correct small offsets in phase or frequency.
The resulting inertia in modifying the clock's parameters hinders quick, large-magnitude clock updates by an attacker.
We already need a TSC frequency $F_{\text{TSC}}$ at protocol~\textbf{A}'s start to measure the duration of the \texttt{FREQ} phase.
To that end, time services have a launch argument to set this frequency: honest nodes can retrieve their OS's accurate measurement of $F_{\text{TSC}}$.

Each \sysname enclave starts with a ``\texttt{TA\_INCONSISTENT}'' clock.
Clock consistency is measured against a maximum absolute offset of $\tau=\mu T$ with respect to the reference clock, with $\mu=15\cdot10^{-6}s/s$ the maximum allowed clock drift rate and $T$ the poll period with the TA, e.g., $T=64s$.
Given the $\mu$ and $T$ values above, if a node has an absolute offset to the TA of less than 960µs at the end of the \texttt{FREQ} phase, it directly proceeds as \texttt{TA\_CONSISTENT} in the \texttt{SYNC} phase. 

While the poll period with respect to the TA is in the order of seconds in the \texttt{FREQ} phase (4s), one or more minutes pass between requests in the \texttt{SYNC} phase: we share with Triad~\cite{fernandezTriadTrustedTimestamps2023} the objective of rare TEE--TA communications compared to within the TEE cluster.
An attacker could try to manipulate the local clock in between checks with the TA: protocols \textbf{B} and \textbf{C} detailed hereafter perform finer-grained checks, avoiding the need for more requests to the TA.
Moreover, adding delays before transmitting the response back to a slowed-down enclave reduces the perceived offset with respect to the TA.
The enclave's current clock can then be consistent with a TA's past clock state instead of the current one.
Protocol \textbf{C} complements \textbf{A} to mitigate this vulnerability.   

\subsection{Local clock monitoring}

With SGX2, \texttt{rdtscp} instructions to read the TSC can be executed without leaving the enclave: the OS or hypervisor cannot tamper with the TSC during a continuous enclave execution.
However, between continuous executions, the TSC can be manipulated by the attacker.
Similarly to Triad~\cite{fernandezTriadTrustedTimestamps2023}, we use an enclave thread to monitor the number of \texttt{INC} instructions that can be executed between a number of TSC increments.
Empirically, high-precision measurements can be obtained for TSC monitoring, e.g., with a standard deviation of 3\texttt{INC}~\cite{bettinger2025trihard}.
TSC monitoring depends on the thread's core frequency. 
Therefore, this core's frequency should be fixed.
Frequency changes restart protocol \textbf{A}: they trigger a new calibration of the measured TSC rate, locally against \texttt{INC} instructions and remotely against the TA's clock speed and time reference.
Furthermore, to reduce the number of interruptions from benign process scheduling by the OS, this core is isolated from the scheduling algorithm and its interruption rate is reduced, at least at honest nodes.

TSC monotonicity is monitored, as well as the magnitude of TSC increments during interruptions.
Checking monotonicity prevents backward jumps in time, but an attacker can still hide an arbitrarily long interruption by setting the TSC to a value close to the one it had at the start of an interruption.
However, interruptions are detected with AEX-Notify~\cite{Constable_Bulck_Cheng_Xiao_Xing_Alexandrovich_Kim_Piessens_Vij_Silberstein_2023}: they break progress in stored TSC updates and therefore timestamp availability by protocol~\textbf{D}, so it is not in the interest of the attacker.
Now, to prevent arbitrary forward jumps in TSC value, our AEX handler checks that the last stored and the current TSC values differ by less than a \emph{panic threshold} of increments.
If the increase in TSC value exceeds the panic threshold, we log a \emph{panic event}.
In practice, panic events can be implemented to abort execution or broadcast to other nodes for accountability.
In the latter case, the node should remain unavailable until it has received $f$ acknowledgments.
The panic threshold can be set using benchmarks on each machine.
Nodes can broadcast their threshold value to others to detect aberrant values. 
Indeed, there should not be arbitrarily long interruptions during honest execution, because the monitoring thread's core is isolated from scheduling and other processes.

Interruptions trigger an AEX event upon resuming enclave execution:
the node's TSC state then switches to ``\texttt{TAINTED}'', i.e., ``possibly manipulated'' by the attacker.
With Triad, low OS interruption rates enable the attacker to prevent its node from trying to communicate with others, leading to arbitrary clock drift while remaining available to serve timestamps~\cite{bettinger2025trihard}.
In this paper, we add events that trigger from within the enclave: if no AEX happens within a given increase in TSC, the enclave ``self-taints'', i.e., switches to a \texttt{TAINTED} TSC state autonomously.
Whether due to an AEX or self-tainting, to switch to an ``\texttt{OK}'' state and be able to serve timestamps, the node's state in protocol \textbf{A} must be \texttt{TA\_CONSISTENT}, and protocol \textbf{C} must succeed, which we now describe.

\subsection{Cluster clock-consistency verification}

Contrary to Triad's approach~\cite{fernandezTriadTrustedTimestamps2023}, communications with enclave peers in \sysname only check for consistency with peer clocks and do not trigger clock adjustments.
When \texttt{TA\_CONSISTENT}, the node tries to verify its clock's consistency by sending NTP request messages to all peers and expects $f$ responses consistent with its own clock.
Only \texttt{TA\_CONSISTENT} peers respond to such requests.
If the requesting enclave is hosted on an honest machine, due to its consistency with the TA, the node cannot be consistent with the $f$ malicious nodes or, if so, those nodes exhibit deviations from honest peers' and the TA's clocks within tolerance bounds, i.e., the attack is weak.
If the requesting enclave's host is malicious, messages coming back from peers may be delayed, invalidating the symmetric network latency assumption and thus the clock offset computation.
As a consequence, as with the TA, that enclave may think itself consistent with honest nodes, but is in reality offset to the past.
To prevent this, responses from peers also contain whether they consider the requesting enclave's clock consistent with theirs.
Note how an enclave on a malicious node can never receive more than $f-1$ malicious peer messages, as it is part of the $f$ malicious nodes.
Upon receiving $f$ consistent messages from peers, the node switches from the \texttt{TAINTED} to the \texttt{OK} state, and becomes available to serve timestamps to client applications.
Communications with peers, hosted in the same cluster, should have low latencies: a peer response is considered usable for consistency checks if there was no AEX or self-tainting event that occurred during the round-trip communication with that peer.
Messages additionally contain a sequence number $s$ to prevent attackers from holding back valid peer messages to play later.
Without sequence numbers, an enclave with a slowed-down clock in a malicious host could be fed old honest messages to erroneously think it is consistent with them and therefore legitimate to serve timestamps.

\subsection{Timestamp service}

Finally, each \sysname node implements an interface for client applications to get the enclave clock's current timestamp.
Timestamps are only returned when the node is in the \texttt{OK} state, i.e., to summarize: its clock is consistent with the TA and $f$ peers, and no AEX nor self-tainting event occurred since the last round of protocol \textbf{C}'s peer consistency check, which was successful.
Additionally, since the AEX-Notify handler runs when the enclave \emph{resumes} execution, even if the state is \texttt{OK} when a client tries to get a timestamp, protocol \textbf{B}'s TSC monitoring thread may be currently interrupted and may have been so for an arbitrarily long time.
To prevent this behavior, the \texttt{get\_timestamp} operation checks not only for the \texttt{OK} state but also that the monitoring thread progresses: it waits for an increase in the stored TSC value while in the \texttt{OK} state.
A client enclave can check that it gets and uses the timestamp within a continuous execution by leveraging AEX-Notify~\cite{Constable_Bulck_Cheng_Xiao_Xing_Alexandrovich_Kim_Piessens_Vij_Silberstein_2023} in its logic as well, creating a transaction.

%% file: content/experimental_protocol.tex
\section{Experimental protocol}\label{sec:experimental_protocol}

We now describe how we evaluate our proposed protocol \sysname's implementation, whose baseline is the public Triad implementation~\cite{Triadcode}. 
For reproducibility, we provide repositories containing the source code and experiments~\cite{TriHaRdcode,Triadcode}.

\subsection{Metrics}

We measure the drift, access latency, and availability of the enclave time services' timestamps for Triad and \sysname. 
First, we periodically get timestamps at each node from its time service enclave and compute the offset to the machine's wall clock when the call returns, i.e., we evaluate drift when timestamps are received or used, not compared to when they were requested.
Timestamp access latency is measured as the number of cycles spent calling the enclave time service's \texttt{get\_timestamp}.
We measure it both for client applications that run in the same enclave as the time service, as well as for non-TEE client applications.
In both cases, we compare time service access latencies to accessing libc's \texttt{timespec\_get}.
Time service enclave availability is the share of time where client applications can get timestamps, e.g., with \sysname, the time spent in the \texttt{OK} state.
This metric is not a success rate (clients wait, possibly indefinitely, for a timestamp): it complements the access latency metric's discrete measurements with a more continuous view of the system's behavior.

For explainability, we also log events (e.g., AEXs, switching to a given state) with respect to the node's wall clock time.
Note that, rarely, in about 1 in 10 experiments, one timestamp given by libc's \texttt{timespec\_get} out of the whole logging timeseries was aberrant, e.g., returning the Unix epoch as a date.
We filter out those outliers in log timestamps.
Time-service enclaves timestamps are never excluded or modified.

\subsection{Deployments}

\begin{figure*}
    \centering
    \begin{subfigure}{0.49\textwidth}
        \centering
        \includegraphics[trim={0mm 3mm 0 0mm}, scale=0.78]{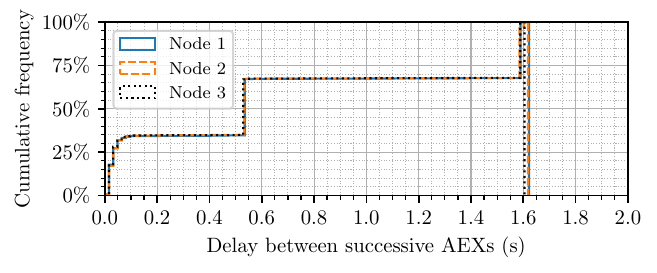}
        \subcaption{Triad-like~\cite{fernandezTriadTrustedTimestamps2023} simulated inter-interruption delay distribution without self-tainting.}
        \label{fig:aexdelaytriad}
    \end{subfigure}
    \begin{subfigure}{0.49\textwidth}
        \centering
        \includegraphics[trim={0mm 3mm 0 0mm}, scale=0.78]{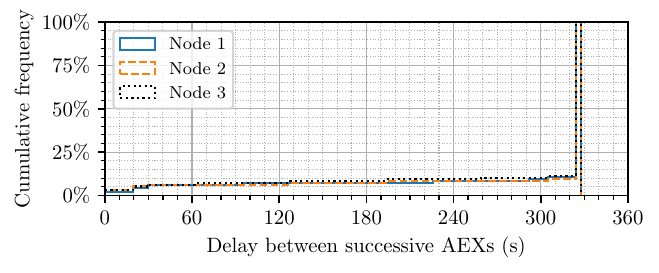}
        \subcaption{TSC monitoring core isolated from most OS interruptions without self-tainting.}
        \label{fig:aexdelaylowinterrupt}
    \end{subfigure}
        \begin{subfigure}{0.49\textwidth}
        \centering
        \includegraphics[trim={0mm 3mm 0 0mm}, scale=0.78]{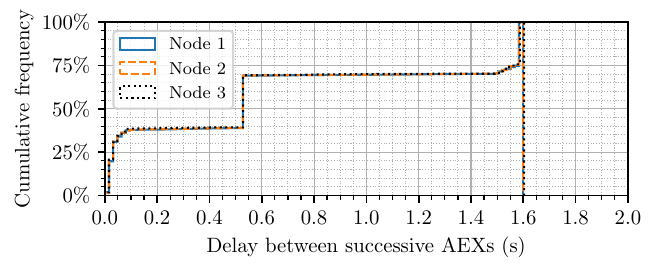}
        \subcaption{Distribution of delays between tainting events using Triad-like~\cite{fernandezTriadTrustedTimestamps2023} interruptions and self-tainting set to 1.5s.\\}
        \label{fig:aexdelaytrihard-staint}
    \end{subfigure}
    \begin{subfigure}{0.49\textwidth}
        \centering
        \includegraphics[trim={0mm 3mm 0 0mm}, scale=0.78]{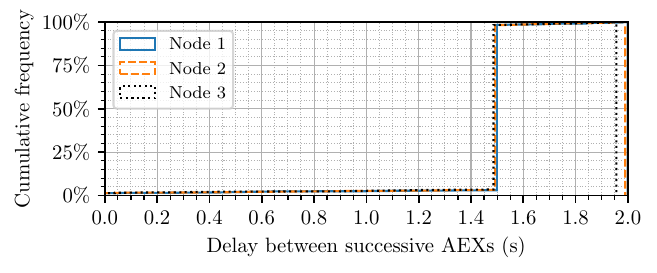}
        \subcaption{Distribution of delays between tainting events using a TSC monitoring core isolated from most OS interruptions and self-tainting set to 1.5s.}
        \label{fig:aexdelaylowinterrupt-staint}
    \end{subfigure}
    \caption{Cumulative distribution of delays between successive Asynchronous Enclave Exits (AEXs) on the TimeStamp Counter (TSC) monitoring enclave thread at each time service. 
    \Cref{fig:aexdelaytriad} is simulated on top of \Cref{fig:aexdelaylowinterrupt}'s environment, by triggering AEXs at the TSC monitoring thread's core, using \texttt{rdmsr} instructions on that core's TSC MSR (Model Specific Register, \texttt{0x10} for TSC).
    \Cref{fig:aexdelaytrihard-staint,fig:aexdelaylowinterrupt-staint} show \sysname-perceived delays in TSC tainting events when using self-tainting set to 1.5s.
    }
    \label{fig:aexdelay}
    \vspace{-0.25cm}
\end{figure*}

In experiments, we deploy \sysname on a single machine or on multiple machines.
In both cases, we simulate a 30ms round-trip network latency between time service enclaves and the TA.
Additionally, we isolate the TSC monitoring thread from the scheduling algorithm and reduce its interruption frequency.
We reproduce inter-interruption delay distributions used on Triad~\cite{bettinger2025trihard}: 
either a ``Triad-like'' distribution (i.e., delays of 10ms, 532ms, and 1.59s, each 1/3 of the time), shown in \Cref{fig:aexdelaytriad}, similar to the original Triad paper~\cite{fernandezTriadTrustedTimestamps2023}; 
or with rare interruptions (\Cref{fig:aexdelaylowinterrupt}), with 90\% every 5.5min, 10\% approximately uniformly distributed between 0 and 5.5min.

The single machine is a 32-core SGX2 Ubuntu 24.04 server with AEX-Notify available.
The TA and three time-service enclaves run on this same server.
For the multi-machine setup, we use four Azure VMs, three of which are 4-core SGX2 VMs, while the fourth 2-core VM hosts the TA.
Note that these Azure VMs are missing a required feature for AEX-Notify, namely EDECCSSA.
Following an AEX and the AEX handler's execution, the leaf function EDECCSSA enables switching contexts back to enclave execution~\cite{WhitePaperAEXEDECCSSA}.
For the multi-machine setup, we simulate the missing AEX-Notify functionality by using the TSC monitoring thread.
If the TSC increments between two \texttt{rdtscp} instructions by more than a threshold (around $10^{3}$ to $10^{4}$ cycles, depending on the machine), we detect it as an AEX event and switch to a \texttt{TAINTED} state.
We cannot use that thread to simultaneously detect both TSC frequency discrepancies and AEX events: as a consequence, we attack the TSC register itself in single machine experiments but not in multi-machine experiments.

\subsection{Implementation and configuration}

For more comparable results, we use the public Triad implementation~\cite{Triadcode} as a base code architecture for \sysname. 
We keep the same interface with client applications, the TSC monitoring and AEX handler code (we add \sysname's panic threshold check), event logging, lower-level encryption, and network communication code.
Enclaves communicate with each other and the TA using AES-GCM-encrypted UDP for Intel SGX~\cite{Li_2020}.
On top of this, we replace Triad's calibration and peer untainting protocols with \sysname protocols \textbf{A}--\textbf{D}.

TSC monitoring rounds each last 500ms. Our self-tainting threshold is set to 1.5s, i.e., it triggers if no AEX occurs during 3 consecutive monitoring rounds.
1.5s is chosen to be close to the largest inter-interruption delay in the Triad-like distribution (\Cref{fig:aexdelaytriad}).  
The panic threshold is set with micro-benchmarks of interruption durations on the hardware used in experiments.
In practice, panic events are triggered when the TSC increases by more than 100µs during an interruption.

\subsection{Attack vectors}

Previous work shows how TSC monitoring can detect local changes in TSC frequency~\cite{fernandezTriadTrustedTimestamps2023,bettinger2025trihard} (due to attacks on the TSC or monitoring core frequency-scaling), so we do not evaluate it in this paper.
Instead, we focus on manipulating the perceived TSC offset and increment rate with respect to the TA's reference time and speed.
In particular, at one node out of the 3, i.e., considering $f=1$, we strategically make forward jumps in the TSC value and we delay some communications with the TA and peer enclaves. 

%% file: content/results.tex
\section{Results}\label{sec:results}

We evaluate \sysname along the following research questions RQA and RQB, detailed in the section with matching letter: 
\begin{itemize}
    \item[\emph{A}:] How available and accurate are \sysname's time service enclaves in single- and multi-machine setups?
    \item[\emph{B}:] How resilient is \sysname to clock manipulation attacks?
\end{itemize}
In all the following figures, Nodes 1, 2, and 3 are always depicted in blue, orange, and black, respectively.
Under attacks, Node 3 is malicious while Nodes 1 and 2 are honest.

\subsection{Fault-free drift and availability}

\begin{figure*}
    \begin{subfigure}{0.49\linewidth}
        \includegraphics[trim={-0.5mm 3mm 0 0mm}, scale=0.78]{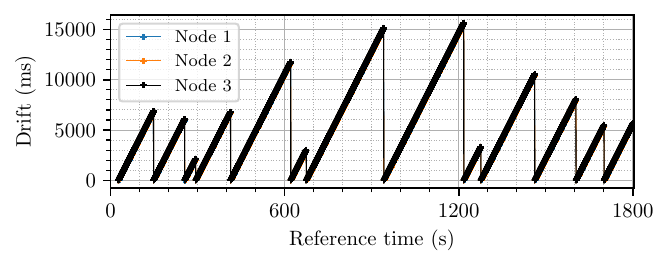}
        \subcaption{Clock drift per Triad node over time with $F_{1}^{\text{calib}}=2901.187\text{MHz}$; $F_{2}^{\text{calib}}=2900.246\text{MHz}$; $F_{3}^{\text{calib}}=2744.152\text{MHz}$.}
        \label{fig:driftff-triad}
    \end{subfigure}
    \begin{subfigure}{0.49\linewidth}
        \includegraphics[trim={-1mm 3mm 0 0mm}, scale=0.78]{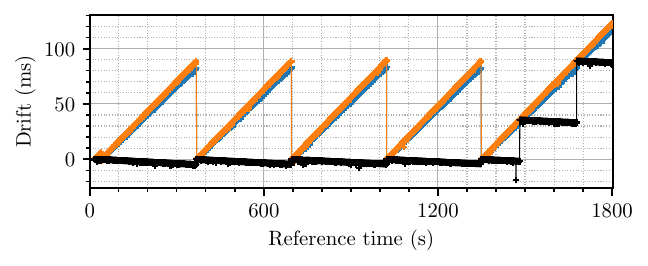}
        \subcaption{Clock drift per Triad node over time with $F_{1}^{\text{calib}}=2899.227\text{MHz}$; $F_{2}^{\text{calib}}=2899.174\text{MHz}$; $F_{3}^{\text{calib}}=2900.003\text{MHz}$.}
        \label{fig:driftfflowinterr-triad}
    \end{subfigure}
    \begin{subfigure}{0.49\linewidth}
        \includegraphics[trim={0mm 3mm 0 0mm}, scale=0.78]{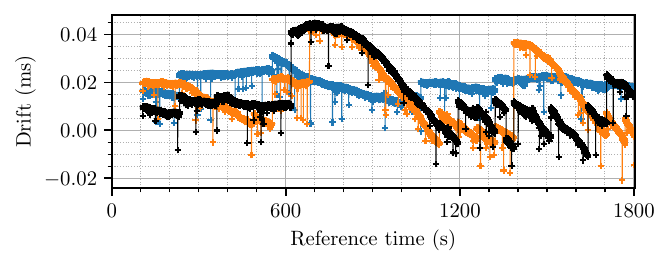}
        \subcaption{Clock drift per \sysname node over time.}
        \label{fig:driftff-trihard}
    \end{subfigure}
    \begin{subfigure}{0.49\linewidth}
        \includegraphics[trim={0mm 3mm 0 0mm}, scale=0.78]{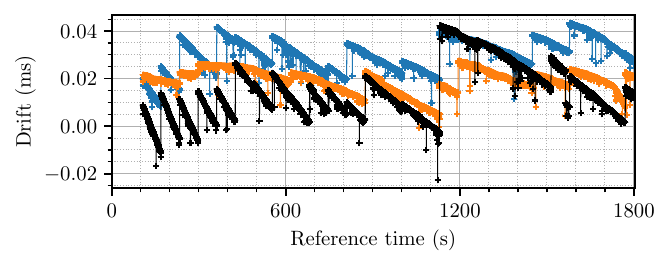}
        \subcaption{Clock drift per \sysname node over time.}
        \label{fig:driftfflowinterr-trihard}
    \end{subfigure}
    \caption{Single-machine long-term fault-free behavior of Triad (top) and \sysname (bottom) nodes.
    The inter-interruption/self-tainting delay distribution for \Cref{fig:ff}.i is illustrated by \Cref{fig:aexdelay}.i.
    OS-measured TSC clock frequency is $F_{\text{TSC}}=2899.999\text{MHz}$.
    }
    \label{fig:ff}
    \vspace{-0.33cm}
\end{figure*}

First, we measure the drift, access latency and availability of timestamps served by protocol~\textbf{D}, in a context without attacks.
\Cref{fig:ff} presents drift compared to the OS clock for Triad and \sysname nodes on the single machine.
For Triad nodes, whether with higher (\Cref{fig:driftff-triad}) or lower interruption rates (\Cref{fig:driftfflowinterr-triad}), all three nodes drift significantly, in the order of hundreds of microseconds per second or more.
Meanwhile, in both cases, \sysname nodes drift within 50µs from the reference, at a maximum instantaneous rate around $\pm0.3$ppm, i.e., within NTP's 15ppm tolerance bound.

Another observation is the independence in drift behavior between \sysname nodes.
On the contrary, Triad's peer untainting protocol forces nodes to jump to the clock with the highest value at each interruption, hence the unified sawtooth pattern in \Cref{fig:driftff-triad}.
The sawtooth pattern itself is due to nodes getting new TA time references when all nodes are interrupted simultaneously.
Between interruptions, their clocks update at their own rates: with rare interruptions in \Cref{fig:driftfflowinterr-triad}, Node 3 drifts negatively, until Node 1 makes it jump to higher clock values, e.g., with significant drift at 1481s and 1677s.

\begin{figure*}
    \begin{subfigure}{0.49\linewidth}
        \includegraphics[trim={-1mm 3mm 0 0mm}, scale=0.78]{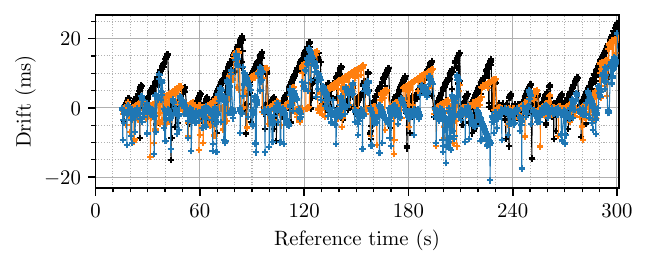}
        \subcaption{Clock drift per Triad node over time ($F_{1}^{\text{calib}}=2795.143\text{MHz}$; $F_{2}^{\text{calib}}=2794.691\text{MHz}$; $F_{3}^{\text{calib}}=2796.026\text{MHz}$).}
        \label{fig:driftff-multi-triad}
    \end{subfigure}
    \begin{subfigure}{0.49\linewidth}
        \includegraphics[trim={3mm 3mm 0 0mm}, scale=0.78]{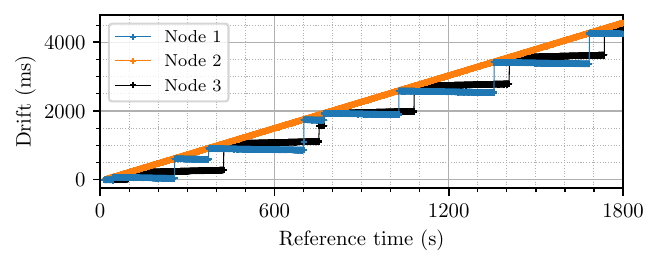}
        \subcaption{Clock drift per Triad node over time ($F_{1}^{\text{calib}}=2793.910\text{MHz}$; $F_{2}^{\text{calib}}=2786.368\text{MHz}$; $F_{3}^{\text{calib}}=2792.892\text{MHz}$).}
        \label{fig:driftfflowinterr-multi-triad}
    \end{subfigure}
    \begin{subfigure}{0.49\linewidth}
        \includegraphics[trim={0mm 3mm 0 0mm}, scale=0.78]{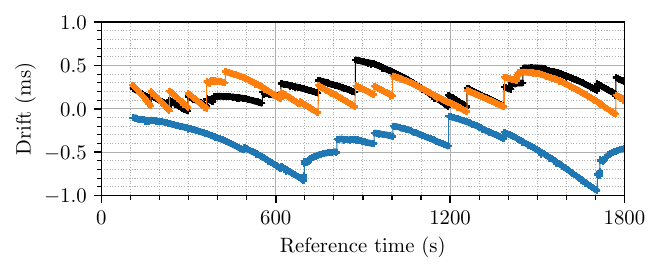}
        \subcaption{Clock drift per \sysname node over time.}
        \label{fig:driftff-multi-trihard}
    \end{subfigure}
    \begin{subfigure}{0.49\linewidth}
        \includegraphics[trim={0mm 3mm 0 0mm}, scale=0.78]{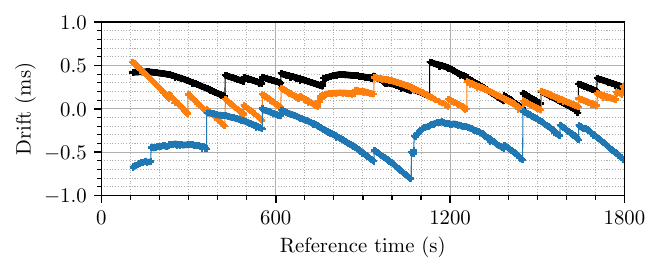}
        \subcaption{Clock drift per \sysname node over time.}
        \label{fig:driftfflowinterr-multi-trihard}
    \end{subfigure}
    \caption{Multi-machine long-term fault-free behavior of Triad (top) and \sysname (bottom) nodes under Triad-like interruptions (left) and in a low AEX environment (right).
    OS-measured TSC clock frequency is $F_{\text{TSC}}=2793.439\text{MHz}$ for all machines.
    Note that in \Cref{fig:driftff-multi-trihard,fig:driftfflowinterr-multi-trihard}, Node 1's OS clock, used as reference for its own enclave time service drift, may have exhibited a slight offset (around -0.5ms) compared to Nodes 2 and 3's OS clocks, hence the negative drift correction, while the drift also looks negative.
    Successful peer consistency checks (with a tolerance of 500µs) between enclave clocks support this argument.
    }
    \label{fig:ff-multi}
    \vspace{-0.33cm}
\end{figure*}

Similar conclusions hold in the multi-machine setup: \sysname nodes stay within 1ms of the reference (\Cref{fig:driftff-multi-trihard,fig:driftfflowinterr-multi-trihard}), while Triad nodes drift by tens of milliseconds (\Cref{fig:driftff-multi-triad}) or seconds (\Cref{fig:driftfflowinterr-multi-triad}).
The highest-drifting Triad clock leads others to jump positively, for an arbitrarily long time.
Indeed, Node 2 in \Cref{fig:driftfflowinterr-multi-triad}, with the lowest calibrated frequency, drifts linearly, while Nodes 1 and 3 regularly jump forward in timestamp, creating a staircase pattern.


Regarding availability, using the same single-machine experiment runs as with drift, Triad is an \texttt{OK} state 99.8\% of the time with Triad-like interruption rates, 99.9\% with low interruption rates.
Note, however, that from previous experiments, drift rises when interruption rates are low.
Instead, \sysname's drift is relatively stable in both cases, and availability remains around 99.98\% and 99.995\% when interruptions are respectively Triad-like and rare.
With rare interruptions and without any self-tainting, \sysname reaches ``6 nines'' of availability. 

Regarding timestamp access latencies, under Triad-like interruptions, when the client application runs within the time service enclave, it takes in the order of 400--4000 TSC cycles (0.14--1.4µs) to get a timestamp, while getting the OS time from outside the enclave takes around 14k--30k cycles (4.8--10µs).
When the client application is outside the enclave instead, it takes 20k--80k cycles (7--28µs) using the time service and 200--2000 cycles (0.07--0.7µs) via the OS.
In other words, interface calls to and from the enclave represent a significant source of latency, hence the lower latencies whenever they are not necessary.
\sysname enclave time service timestamps for colocated enclave client applications and OS timestamps for traditional client applications show similar orders of magnitude in access latencies.
Triad exhibits similar access latencies for enclave clients with 300--3000 cycles, but 0.1M--10M cycles (34µs--3.4ms) for traditional clients. 

\subsection{Attacking \sysname}

Now, we assess \sysname's resilience against attacks: first by leveraging the configurable initial $F_{\text{TSC}}$ at node start, then by extending this attack using strategically manipulated TSC values to avoid detection, in the single-machine setup.

\subsubsection{Misconfigured initial TSC frequency (protocol \textbf{A})}

We let each node set an initial TSC frequency $F_{\text{TSC}}$, at the startup of the enclave time service.
This frequency determines the duration of the \texttt{FREQ} phase: 
it is used in timers for communications with the TA to perform the frequency estimation used in subsequent phases.
We enable this behavior because a predefined frequency could help honest nodes get more accurate $F_{\text{TSC}}$ estimations, but it also opens an attack vector for malicious ones.
Honest nodes retrieve the correct TSC frequency $F_{\text{TSC}}=2899.999\text{MHz}$ from OS startup logs, while malicious nodes may input a frequency with an offset $o$ compared to $F_{\text{TSC}}$.
Previously~\cite{bettinger2025trihard}, with Triad, malicious nodes were shown to be capable of manipulating calibration with the TA to set an arbitrary $F_{\text{TSC}}$ value.

Now, with \sysname, we observe the impact of the offset $o$ on the malicious node's ability to participate in subsequent sub-protocols and whether its estimated frequency through the \texttt{FREQ} phase is significantly different from honest nodes.
We perform a dichotomic search of cut-off $o$ values where the node's behavior in the protocol changes.
With $o>0$, the node's clock advances slower and the TSC frequency is correctly estimated in all runs.
Indeed, due to the lower clock speed, the \texttt{FREQ} phase takes more time (which is beneficial to accurate estimation).
Faster node clocks instead shorten the \texttt{FREQ} phase duration: 
While a node with a slower clock can be offset forward in time towards the TA, faster clocks cannot be offset back in time.
Nodes with a misconfigured frequency by an offset $o\in\rbrack-170;197\rbrack$MHz become \texttt{TA\_CONSISTENT} at the end of the \texttt{FREQ} phase. 
With $o<-170$MHz or $o>197$MHz, nodes fail the TA consistency check.
In all cases except very fast misconfigured clocks (e.g., $F_{\text{TSC}}+o<2\text{GHz}$), the frequency estimated during the \texttt{FREQ} phase is set within a range of $\pm10\text{Hz}$, both on honest and malicious nodes.

Protocols \textbf{A} and \textbf{B} themselves do not protect against forward jumps in TSC offset: 
a malicious node with a slow clock may offset the TSC before the request to the TA and then between the request and response, to make it seem like time passes normally.
Consequently, the node can become \texttt{TA\_CONSISTENT} through the first check and proceed to other sub-protocols.
This is a known attack vector, whose protection by protocol \textbf{C}'s peer consistency checks coupled with \textbf{B}'s clock tainting upon AEXs we evaluate now.

\subsubsection{Attacking consistency checks}

To serve malicious timestamps in protocol~\textbf{D}, malicious nodes must first pass protocol~\textbf{C}.
A key constraint is that both the enclave in the malicious node and the $f$ peers must consider each other's clocks as consistent simultaneously.
Similarly to before in protocol~\textbf{A}, an attacker can increase the TSC just before sending an untainting request to peers.
The main difference with the same attack in protocol~\textbf{A} is that, in \textbf{C}'s short-distance communications with peers, there must not be any AEX during the round-trip.
In other words, the attacker cannot increase the TSC between the request and response as before, because such an update causes an enclave exit~\cite{intelManual}.
Moreover, because consistency checks are bidirectional, the enclave must also serve correct timestamps to at least one honest node (in the following, Node 2), any time that honest node sends an untainting request, to have $f$ positive peer checks.
Finally, a malicious node cannot offset the TSC by more than the AEX handler's panic threshold minus the time it takes to perform the update.
To bypass the panic threshold, the attacker may perform multiple incremental updates below the threshold, letting the AEX handler run between two updates.
However, frequent AEXs to adapt the TSC may reduce availability in protocol~\textbf{D}.

Therefore, we evaluate here how slow a malicious node's clock can be without being detected by the AEX handler or rejected by an honest peer, as well as that malicious node's timestamp drift and availability.
For easier reproducibility, malicious TSC updates are instrumented directly in the enclave code, instead of an attacker monitoring the TSC and messages from outside the enclave.
The malicious code to offset the TSC and the corresponding AEX handler simulation take around the same duration as the honest AEX handler (80--100µs in both cases), so availability results are representative.

\begin{figure}
    \begin{subfigure}{\linewidth}
        \includegraphics[trim={2.5mm 3mm 0 0mm}, scale=0.78]{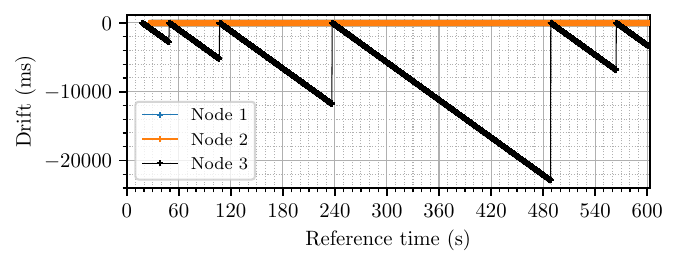}
        \subcaption{Clock drift of Triad nodes under an F+ attack on Node 3 with $F_{3}^{\text{calib}}=3190.041\text{MHz}$; $F_{1}^{\text{calib}}=2900.131\text{MHz}$; $F_{2}^{\text{calib}}=2900.530\text{MHz}$.}
        \label{fig:driftfpone}
    \end{subfigure}
    \begin{subfigure}{\linewidth}
        \includegraphics[trim={0mm 3mm 0 0mm}, scale=0.78]{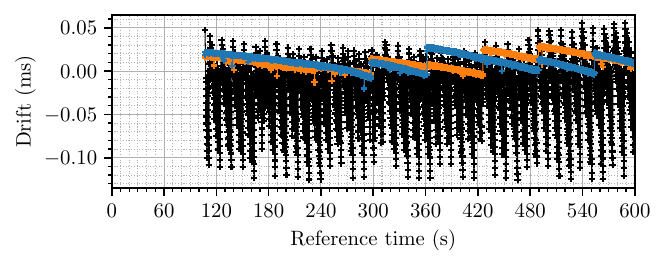}
        \subcaption{Clock drift per \sysname node over time under a forward TSC jump attack on Node 3 (with frequency offset $o=+0.2\text{MHz}$).}
        \label{fig:drifttscjump-trihard}
    \end{subfigure}
    \caption{Attacks slowing down Node 3, in the same-machine setup. 
    Node 3 is in a low AEX environment, while Nodes 1 and 2 experience Triad-like AEXs.}
    \label{fig:tscjump}
    \vspace{-0.33cm}
\end{figure}

Due to the different protocols used in Triad and \sysname, we evaluate attack vectors intrinsic to each, but working towards the same system impact.
\Cref{fig:tscjump} compares the ``F+'' attack on Triad~\cite{bettinger2025trihard}, which overestimates $F_{\text{TSC}}$ (and so slows the clock), to \sysname's misconfigured slow clock and strategic TSC forward jumps.
Honest nodes 1 and 2 experience Triad-like interruptions (\Cref{fig:aexdelaytriad} for Triad, \Cref{fig:aexdelaytrihard-staint} for \sysname).
In both cases, Node 3's TSC monitoring thread is isolated from interruptions by the attacker (\Cref{fig:aexdelaylowinterrupt} for Triad, \Cref{fig:aexdelaylowinterrupt-staint} for \sysname), and no detection mechanism is triggered by malicious behavior.
Ultimately, however, attack power is higher with Triad than \sysname.
With \sysname, timestamp drift at Node 3 remains within 130µs, while Triad's Node 3 drifts negatively at -93ms/s.
Indeed, whereas Triad's Node 3 will not check its clock with peers until a rare AEX, \sysname's Node 3 self-tainting every 1.5s regularly requires protocol~\textbf{C}'s successful consistency checks.
As shown in \Cref{fig:aexdelaytrihard-staint,fig:aexdelaylowinterrupt-staint}, whether with high or low interrupt rates, duration stayed in the \texttt{OK} state is restricted to up to 2s, whereas it could reach 5.5 minutes without self-tainting (\Cref{fig:aexdelaylowinterrupt}). 
Moreover, one honest node must perceive a consistent clock at Node 3 before declaring it so: \sysname's Node 3 must jump in TSC value before responding to the honest node, without exceeding the panic threshold (100µs).
The consequence is a maximum frequency offset of $o=+0.2\text{MHz}$ to avoid panic events with \sysname using single TSC jumps, while $o=+291\text{MHz}$ for Triad's Node 3.
Additionally, while honest nodes still experience 99.98\% availability like in the case without attacks, at the attacked node, interruptions due to single forward TSC jumps reduce its availability to 60\% and produces access latency outliers in the order of 0.1--1s. 
Allowing multiple TSC jumps enables higher frequency offsets but reduces availability even more.
In other words, the attack has a low impact on drift (compared to the reception time) and reduces the quality-of-service (i.e., with respect to availability and access latencies).
Recall that related work's threat models~\cite{fernandezTriadTrustedTimestamps2023,hamidyT3EPracticalSolution2023a} also assume attackers aim to not affect the latter during their attacks.
Finally, considering faster clocks, we do not show figures for the sake of space but we describe results: while Triad's update policy enables drift in the order of hundreds of milliseconds per second or more (with a behavior similar to \Cref{fig:driftff-triad}), 
in experiments, \sysname's TSC monotonicity and peer consistency checks always lead the malicious node to become unavailable after it exceeds the consistency threshold, i.e, the attack is detected and prevented.

\subsection{Summary}

To conclude and to answer RQA, \sysname upholds an absolute drift offset within 1ms, with honest clock speeds calibrated within 0.3ppm (0.3µs/s).
Meanwhile, Triad nodes can drift by multiple milliseconds per second and reach multi-second offsets.
Moreover, regarding RQB, \sysname's frequent bidirectional peer consistency checks and its TSC increment's panic threshold drastically reduce the attack power, e.g., reducing the maximum undetected clock speed offset by more than 3 orders of magnitude compared to Triad.
Self-tainting also provides a strong protection against malicious hosts isolating their nodes from peers to increase drift.

%% file: content/conclusion.tex
\section{Conclusion}\label{sec:conclusion}

Providing  trusted time to TEE-based systems is an important requirement for enabling reliable confidential computing.
In this paper, we presented \sysname, a protocol that, upon detecting interruptions of a TEE enclave, verifies the TEE clock's consistency with peers in a Byzantine-resilient manner to prevent attacks on the clock's speed or offset.
Using deployments on single and multiple SGX-enabled machines, with \sysname, we empirically show high resilience against attacks under which related work like Triad remains vulnerable.

%% file: content/bibliography.tex
%
%
%
\bibliographystyle{IEEEtran}
\bibliography{imports/bibliography}
%